\documentclass[useAMS,usenatbib]{mn2e}
\usepackage{color,graphicx}
\usepackage{multirow}
\def\aj{AJ}             	
\def\araa{ARA\&A}       	
\def\apj{ApJ}           	
\def\apjl{ApJ}          	
\def\aap{A\&A}          	
\def\mnras{MNRAS}       	

\def\gsim{\hspace{0.3em}\raisebox{0.4ex}{$>$}\hspace{-0.75em}\raisebox{-.7ex}{$\sim$}\hspace{0.3em}}
\def\lsim{\hspace{0.3em}\raisebox{0.4ex}{$<$}\hspace{-0.75em}\raisebox{-.7ex}{$\sim$}\hspace{0.3em}}

\title[The effect of ISM model]{Clumpy galaxies in cosmological simulations: The effect of ISM model}
\author[S. Inoue \& N. Yoshida]
{\parbox[t]{\textwidth} 
{Shigeki Inoue$^{1,2,3,4}$\thanks{E-mail: shigeki.inoue@nao.ac.jp} \& Naoki Yoshida$^{1,2}$}
\\ \\
$^{1}$Kavli Institute for the Physics and Mathematics of the Universe (WPI), UTIAS, The University of Tokyo, Chiba 277-8583, Japan\\
$^{2}$Department of Physics, School of Science, The University of Tokyo, Bunkyo, Tokyo 113-0033, Japan\\
$^{3}$Center for Computational Sciences, University of Tsukuba, Ten-nodai, 1-1-1 Tsukuba, Ibaraki 305-8577, Japan\\
$^{4}$Chile Observatory, National Astronomical Observatory of Japan, Mitaka, Tokyo 181-8588, Japan
}

\begin{document}

\pagerange{\pageref{firstpage}--\pageref{lastpage}} \pubyear{2014}

\maketitle

\label{firstpage}

\begin{abstract}
We study influence by models of inter-stellar medium (ISM) on properties of galaxies in cosmological simulations. We examine three models widely used in previous studies. The ISM models impose different equations of state on dense gas. Using zoom-in simulations, we demonstrate that switching the ISM models can control formation of giant clumps in massive discs at redshifts $z\sim1$--$2$ while their initial conditions and the other settings such as stellar feedback are unchanged. Thus, not only feedback but ISM models can also be responsible for clumpy morphologies of simulated galaxies. We find, however, that changing the ISM models hardly affects global properties of galaxies, such as the total stellar and gas masses, star formation rate, metallicity and stellar angular momentum, irrespective of the significant difference of clumpiness; namely the ISM models only change clumpiness of discs. In addition, our approach provides a test to investigate impact by clump formation on the evolution of disc galaxies using the same initial conditions and feedback. We find that clump formation does not significantly alter the properties of galaxies and therefore could not be the causes of starburst or quenching. 

\end{abstract}


\section{Introduction}
\label{Intro}
Recent observations suggest that star-forming galaxies tend to have clumpy morphologies, and the structures which are often referred to as `giant clumps' are detected with star-forming lights such as H$\alpha$ and ultraviolet (UV) emission. Number fractions of these clumpy galaxies increase with redshifts $z$ up to $\sim50$ per cent at $z\simeq2$--$3$, depending on their masses \citep{tkt:13II,mkt:14,gfb:14,sok:16,bmo:17}. Although a substantial fraction of such clumpy galaxies can be ongoing mergers \citep[e.g.,][]{w:06,fgb:09,p:10,r:17}, giant clumps are considered to reside in disc structures of galaxies since the majority of star-forming galaxies at $z\simeq1$--$2$ are observed to be rotation-supported systems \citep[e.g.][]{w:15}. Accordingly, formation of giant clumps in these high-redshift galaxies is thought to be driven by dynamical instabilities such as Toomre instability \citep[e.g.][]{n:98,n:99,dsc:09}, fragmentation of spiral arms \citep{iy:18b,iy:18} and other non-linear processes \citep{idm:16}. The high-redshift clumpy galaxies may be progenitors of low-redshift spiral galaxies \citep[e.g.][]{SN:93,n:98,n:99}. It is important to study the nature of star-forming galaxies hosting giant clumps since giant clumps can be relevant to formation processes and dynamical properties of galactic structures such as discs, bulges, dark matter (DM) haloes and halo objects \citep[e.g.][]{eev:05,ebe:08,ee:06,bee:07,es:13,is:11,is:12,is:14,nb:15,c:19}.

Cosmological simulations are often used to study formation and evolution of galaxies. For clumpy galaxies, however, results of previous studies are divergent. Cosmological zoom-in simulations performed by \citet[][VELA simulations]{cdb:10,cdm:11,ckk:14} demonstrate that gas-rich galaxies can form giant clumps, and some clumps are massive enough to be gravitationally bound structures and finally fall into the galactic centre in their orbital time-scales. Such clumps are typically long-lived \citep{mdc:13}. In new versions of VELA simulations where the effect of radiation pressure is also included in addition to supernovae (SNe) and runaway stars\footnote{Thirty per cent of the newly formed stellar particles are assigned velocity kick of $\sim10~{\rm km~ s^{-1}}$. They can migrate $\sim100~{\rm pc}$ away from dense regions and explode as SNe in low-density regions where cooling times are long \citep{cdb:10}.}, \citet{mgm:14} and \citet{mdc:17} show that massive clumps $\gsim10^{8}~{\rm M_\odot}$ form and survive until merging with central bulges, although the numbers of low-mass clumps are reduced significantly. \citet{g:12} take into account effects of strong winds from clumps with their parametric method in their simulations and show that a clump cannot grow massive and is disrupted by galactic tide soon after its formation; they propose that giant clumps are short-lived and transient structures. \citet[][NIHAO simulations]{bmo:17} perform mock observations for their cosmological simulations using post-process radiation transfer. They find that the clumps are bright only in star-forming lights and hardly seen in stellar and gas density distributions; the `apparent' clumps are short-lived structures, and do not migrate to the galactic centres. \citet[][FIRE simulations]{okh:17} also show that most of clumps have high virial parameters with $\alpha\gsim1$, indicating only marginally bound or unbound states. Their clumps disappear in $\sim20~{\rm Myr}$ after their formation. The variety of the findings in the above studies is generally attributed to differences in their stellar feedback processes such as SNe and radiation from massive stars. Strong feedback can prevent a dense gas cloud from collapsing and/or further accretion of surrounding gas once stars form within it. Previous studies suggest that formation of clumps can be used as a testbed of feedback processes in simulations \citep[e.g.][]{tms:14,mtl:16}. 

In this study, we examine the effect of ISM models in cosmological simulations. In particular, we investigate the influence of the models on `clumpiness' and physical properties of simulated galaxies. ISM models enforce simplified equations of state (EOS) on dense gas without self-consistently solving cooling and heating effects. ISM models are often designed to take into account feedback effects such as SNe with analytical prescriptions (see Section \ref{twophasemodel}). In this sense, ISM model is an alternative way to implement the feedback effects in an approximated manner. Previous studies using simulations with isolated disc models have shown that EOS of gas significantly affects clumpiness of discs and fragmentation of spiral arms \citep[e.g.][]{sd:08,b:11}. Generally, a softer EOS leads discs to have more clumpy morphologies. However, using a different ISM model may change global properties of galaxies, such as stellar and gas masses, star formation rates (SFRs), metallicities and disc sizes. These properties are often used to calibrate sub-resolution physics by matching simulations with observations. It is naively expected, therefore, that we cannot change ISM models arbitrarily. 

We use zoom-in cosmological simulations to show that difference of ISM models can largely impact on clumpiness of galaxies while their global properties are almost unchanged. In Section \ref{sims}, we describe settings of our simulations and details of the ISM models we examine. Our simulation results are shown in Section \ref{result}, and we discuss the results in Section \ref{discussion}. In Section \ref{conclusions}, we present our conclusions and summary.

\section{Cosmological simulations}
\label{sims}
We use the moving-mesh/$N$-body code {\sc Arepo} \citep{arepo} to perform cosmological simulations. Our simulations take into account various sub-resolution physics responsible for formation and evolution of galaxies, as described in Section \ref{physicalmodel}. Throughout the paper, our adopted cosmological parameters are $\Omega_{\rm m} = 0.307$, $\Omega_{\rm b} = 0.048$ and $\Omega_{\rm \Lambda} = 0.693$ with a Hubble constant of $H_0 = 100h~{\rm km~s^{-1}~Mpc^{-1}}$, where $h = 0.68$ \citep{Plank14}. Settings and physical parameters of our simulations are basically the same as those in the Auriga simulation \citep{ggm:17} that also uses {\sc Arepo}, except initial conditions (see Section \ref{ICs}). We also test different ISM models (see Section \ref{ISMmodel}). Details of the code and physical models considered in our simulations have been presented in \citet[][and references therein]{ggm:17}.

\subsection{Initial conditions}
\label{ICs}
Initial conditions of our simulations are created with a publicly open code {\sc MUSIC} \citep{ha:11}. The simulation box has a comoving side length of $100~{\rm Mpc}$. First, we perform an $N$-body run excluding baryons with a coarse resolution until redshift $z=1$. We then select five haloes of $M_{200}=10^{12}$--$10^{13}~{\rm M_\odot}$ from those detected with a friend-of-friend algorithm \citep{def:85}. Our halo selection is just for sampling haloes within this mass range and does not consider their merger histories nor other properties such as a degree of isolation from other haloes. 

Second, we re-run the simulation including baryons. An ellipsoidal region enclosing all particles within $2R_{200}|_{z=1}$ of a target halo is resolved with higher resolutions, where a DM particle and a gas cell have mass resolutions with $m_{\rm DM}=5.8\times10^5~{\rm M_\odot}$ and $\tilde{m}_{\rm g}=9.0\times10^4~{\rm M_\odot}$, respectively. The zoom simulations are evolved until $z=1$. The moving-mesh code operates mesh regulations such as motions of gas cells, refinement and derefinement so that each gas cell keeps its mass $m_{\rm g}\simeq\tilde{m}_{\rm g}$ within a factor of $2$ throughout the runs. Stellar particles are created with an initial mass-resolution $m_{\rm s}\simeq\tilde{m}_{\rm g}$ within a factor of $2$. In the zoom simulations, the comoving gravitational softening length of a high-resolution DM particle is set to $\epsilon_{\rm DM}=0.74~{\rm kpc}$ with the upper limit of $\epsilon_{\rm DM,max}=0.37~{\rm kpc}$ (physical), and that of a stellar particle scales with its mass as $\epsilon_{\rm s}=\epsilon_{\rm DM}(m_{\rm s}/m_{\rm DM})^{1/3}$ with the upper limit of $\epsilon_{\rm s,max}=0.37~{\rm kpc}$ (physical). The softening length of a gas cell varies with its cell volume $V_{\rm cell}$ as $\epsilon_{\rm g}=2.8[3V_{\rm cell}/(4\pi)]^{1/3}$ with the lower limit of $\epsilon_{\rm g,min}=074~{\rm kpc}$ (comoving); however the size of a gas cell is allowed to become smaller than $\epsilon_{\rm g,min}$ in dense regions with converging flows. Magnetic fields are initially oriented to the same direction and uniform with $10^{-14}~{\rm G}$ (comoving); however the magnetic fields little affect formation histories or morphologies of galaxies \citep{pgg:17}.

\subsection{Sub-resolution physical models}
\label{physicalmodel}
\begin{figure*}
  \includegraphics[bb=0 0 1697 757, width=\hsize]{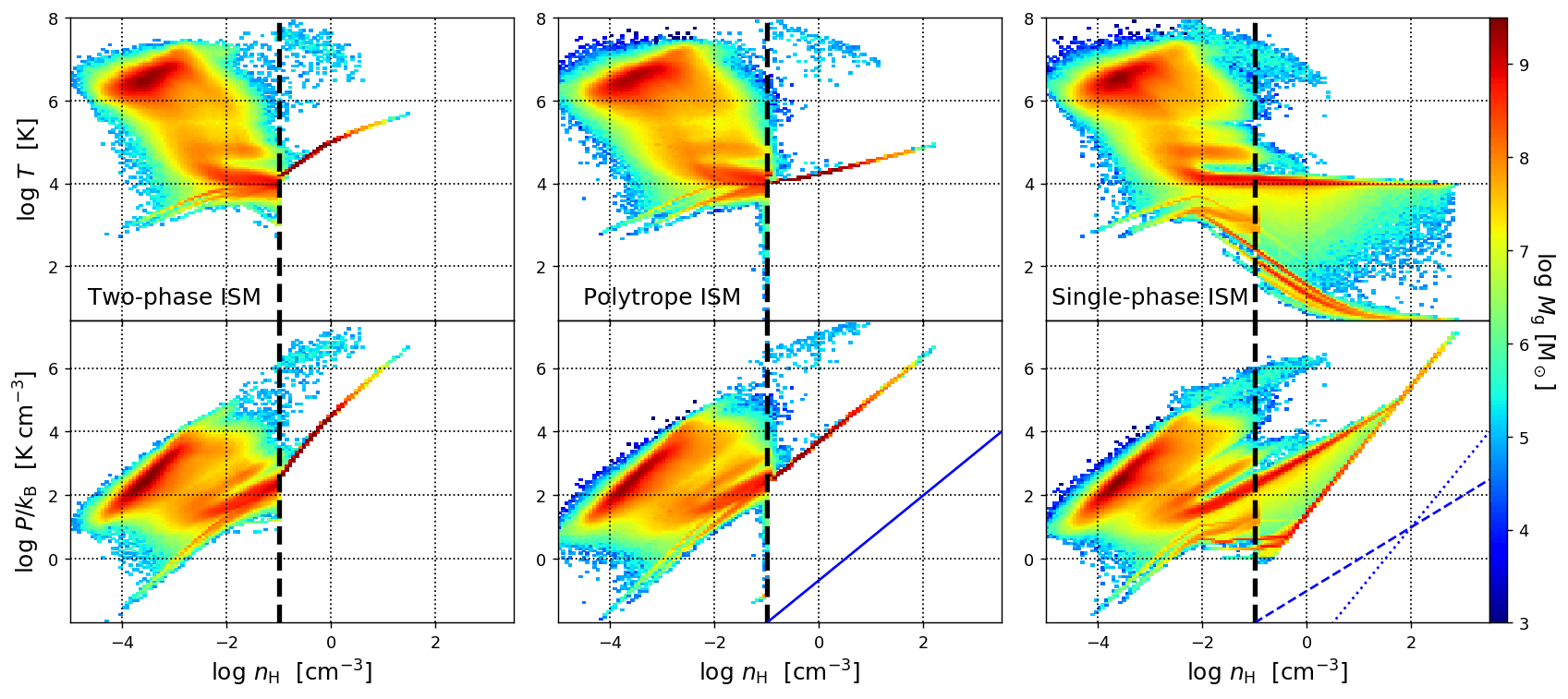}
  \caption{Mass distributions in density--temperature (top) and density--pressure (bottom) diagrams of gas within $R_{200}$ in the run of Halo 1 at $z=1.53$. The colour code indicates gas mass within each bin with $(\Delta\log n_{\rm H},\Delta\log T)=(0.066,0.057)$ and $(\Delta\log n_{\rm H},\Delta\log P)=(0.066,0.074)$ in the top and bottom panels. Left, centre and right sets of panels correspond to the runs with the two-phase, polytrope and single-phase ISM models. The blue solid line in the bottom centre panel delineates the slope of $P\propto n_{\rm H}^{4/3}$ (see Section \ref{polytropemodel}), and the blue dashed and dotted lines in the bottom right panel indicate those of $P\propto n_{\rm H}$ (isothermal) and $P\propto n_{\rm H}^{2}$ (equation \ref{TLC}, see Section \ref{singlephasemodel}). The vertical dashed lines correspond to the star-formation criterion $n_{\rm H,th}=0.13~{\rm cm^{-3}}$ in the simulations, which is set equal to the thresholds above which the two-phase and polytrope ISM models are adopted: $n_{\rm H,th}=\rho_{\rm EOS}$.}
  \label{PhaseDiagrams}
\end{figure*}

In our simulations, dense gas with $\rho_{\rm g}>\rho_{\rm th}$ is considered to be capable of forming stars. Such star-forming gas can create stellar particles according to a stochastic model of star formation, in which an SFR of the gas cell is proportional to its star-forming mass $xm_{\rm g}$ and given as $SFR=0.079 xm_{\rm g}/t_{\rm SF}$ where $m_{\rm g}$ is a mass of the cell, $t_{\rm SF}\equiv(G\rho_{\rm g})^{-1/2}$, and $x$ represents a mass fraction of cold gas which is responsible for star formation. The factor $x$ is computed in ISM models (see Section \ref{ISMmodel}). The star-formation threshold is set to $\rho_{\rm th}=4.50\times10^6~{\rm M_\odot~kpc^{-3}}$ irrespective of temperature or other conditions, which corresponds to $n_{\rm H,th}=0.13~{\rm cm^{-3}}$ in units of the atomic hydrogen number density\footnote{In this study, we denote gas density as $\rho_{\rm g}$ and $n_{\rm H}$ in units of solar mass per cubic kilo-parsec and atomic hydrogen number per cubic centimeter, respectively.} (the vertical dashed lines in Fig. \ref{PhaseDiagrams}).

Each stellar particle is represented with a single stellar population given by the initial mass function (IMF) of \citet{c:05}. The amounts of metal elements produced by stars are computed with tabulated yields: \citet{pcb:98} for asymptotic giant branch (AGB) stars: \citet{k:10} for core-collapse SNe, and \citet{t:03} and \citet{thr:04} for type-Ia SNe. Type-II SNe are assumed to take place immediately after the star formation in the mass range of $8$--$100~{\rm M_\odot}$. Type-II SNe are assumed to release the energy of $1.7\times10^{51}~{\rm erg}$ per SN, and its feedback effect within a star-forming gas cell is modelled by ejecting a wind particle in a random direction \citep{sh:03} with a velocity of $3.46\sigma_{\rm DM}$, where $\sigma_{\rm DM}$ is the one-dimensional velocity dispersion computed from the nearest 64 DM particles \citep[e.g.][]{ofj:10}. The wind particle carries forty per cent of the heavy elements that have been synthesised and expelled by the SNe, and the rest is immediately distributed to nearby gas cells when the star formation occurs \citep{vgs:13}. An orbit of the wind particle is computed only gravitationally, i.e. the hydrodynamical forces are ignored explicitly. The wind particle travels until the maximum travel time or reaching a low-density gas cell with $\rho_{\rm g}<0.05\rho_{\rm th}$. The wind particle finally deposits its mass, metals, momentum and energy into the gas cell in which it is located. Type-Ia SNe and AGB stars eject mass and metals into nearby gas cells every time-step. Our simulations also include creation and feedback of black holes \citep{sdh:05}.

\subsection{ISM models}
\label{ISMmodel}

Basically, gas in our simulations cools by radiative and metal-line cooling with self-shielding corrections computed using {\sc CLOUDY} \citep{cloudy}. Heating by spatially uniform UV background fields is also included, which completes reionization at redshift $z\sim6$ \citep{flz:09}. The temperature and pressure of dense gas are computed in different manners depending on ISM models (see below). We describe details of the ISM models in Sections from \ref{twophasemodel} to \ref{singlephasemodel}. The other physical models and the related parameters are unchanged. 

We examine three ISM models, which are referred to as two-phase, polytrope and single-phase models in this paper. The two-phase ISM model is motivated to take into account the effect of unresolved star formation as sub-resolution physics. Therefore, this model is expected to be realistic if the sub-resolution model is accurate. The polytrope ISM model is not physically motivated but designed to prevent artificial fragmentation of dense gas. Since the EOS of the polytrope model is softer than but similar to that of the two-phase model, the characteristics of the two-phase model is also expected to be somewhat reflected in the polytrope model. The single-phase ISM model self-consistently computes temperature evolution of dense gas from the given cooling/heating rates. Star formation in this model is only represented by creation of stellar particles; therefore this model lacks the effect of star formation and feedback in sub-resolution scales. However, the sub-resolution model would no longer be necessary if resolutions of simulations are high enough to sufficiently capture star formation. In this case, the single-phase model can be expected to achieve realistic representation of dense gas. Since each of these ISM models has its advantage and disadvantage, it is unclear which ISM model is the best for cosmological simulations. Hence, we discuss how changing the ISM models influences disc galaxies in the zoom-in cosmological simulations from the viewpoint of giant clump formation.  



\subsubsection{Two-phase ISM model}
\label{twophasemodel}

The ISM model proposed by \citet{ykk:97} assumes that a single gas cell with $\rho_{\rm g}>\rho_{\rm EOS}$ consists of hot- and cold-phase gas \citep[see also][]{sh:03}; therefore we refer to this model as `two-phase ISM model'. The cold-phase gas hosts \textit{unresolved} stars according to the same IMF assumed in the star formation model (Section \ref{physicalmodel}), and stars more massive than $8~{\rm M_\odot}$ are assumed to instantaneously explode as type-II SNe. The mass fraction of such massive stars is computed to be $\beta=0.23$ in the case of the Chabrier IMF, and the number of type-II SNe is $n_{\rm SN}=0.012~{\rm M_\odot^{-1}}$ in the total mass of new stars formed with the IMF. This means that the average mass of the stars capable of triggering type-II SNe is $m_{\rm SN}=19.1~{\rm M_\odot}$. Although the two-phase ISM model considers the same amount of SFR described in Section \ref{physicalmodel}, the mass fraction $1-\beta$ of the newly formed stars are `unresolved'. Stellar particles are not generated by such unresolved star formation, but the model effectively incorporates the SN feedback caused by the massive stars accounting for the mass fraction $\beta$.

Gas released by the SNe joins the hot phase, and a certain amount of the cold-phase gas is evaporated to the hot phase by SN heating. The amount of the evaporated gas is assumed to be proportional to the mass of the SNe, and the coefficient is expected to be $A\propto\rho_{\rm g}^{-4/5}$ \citep{mo:77}.\footnote{The evaporation factor is normalised to be $A=573$ at $\rho_{\rm g}=n_{\rm th}=0.13~{\rm cm^{-3}}$ in our simulations.} Meanwhile, some amount of hot gas is transferred to the cold phase by thermal instability due to cooling. 

From the three physical processes mentioned above, i.e. SN ejection, evaporation and thermal instability, analytical description of thermal evolution of the two phases can be obtained with only a few parameters such as an SN temperature $T_{\rm SN}=5.73\times10^7~{\rm K}$ and a temperature of the cold phase $T_{\rm c}=10^3~{\rm K}$. These temperatures correspond to the specific energies of $u_{\rm SN}=1.16\times10^{49}~{\rm erg~M_\odot^{-1}}$ and $u_{\rm c}=2.03\times10^{44}~{\rm erg~M_\odot^{-1}}$. The SN feedback in this ISM model is thus estimated to be $u_{\rm SN}m_{\rm SN}=2.22\times10^{50}~{\rm erg}$ per SN, which is $7.7$ times weaker than that implemented using the wind particle model described in Section \ref{physicalmodel}. As shown in the analysis of \citet{sh:03}, the two phases in this model are expected to quickly reach an equilibrium state where star formation is self-regulated. Namely, the evaporation decreases an amount of cold gas when star formation is too active, whereas the thermal instability driven fuels star formation by turning hot gas into cold gas when a mass of hot gas is too large. In this equilibrium state, thermal energy of the hot-phase gas $u_{\rm h}$ is maintained at
\begin{equation}
u_{\rm h}=\frac{u_{\rm SN}}{A+1}+u_{\rm c},
\label{u_hot}
\end{equation}
corresponding to $T_{\rm h}\sim10^6~{\rm K}$. Note that $u_{\rm h}$ in equation (\ref{u_hot}) is a function only of density since the evaporation factor $A$ depends only on $\rho_{\rm g}$. In the self-regulated regime, the effective pressure of the medium, 
\begin{equation}
P_{\rm eff}=(1-\gamma)\left(\rho_{\rm c}u_{\rm c}+\rho_{\rm h}u_{\rm h}\right),
\label{p_eff}
\end{equation}
can be considered to be constant in time, where the ratio of specific heats $\gamma=5/3$, and $\rho_{\rm c}$ and $\rho_{\rm h}$ are masses of cold- and hot-phase gas in the cell. This condition gives
\begin{equation}
\frac{\rho_{\rm c}}{t_{\rm SF}}=\frac{t_{\rm SF}\Lambda_{\rm net}(\rho_{\rm g},u_{\rm h})}{\rho_{\rm g}\left[\beta u_{\rm SN}-\left(1-\beta\right)u_{\rm c}\right]}\frac{\rho_{\rm h}^2}{\rho_{\rm g}^2},
\end{equation}
where $\Lambda_{\rm net}$ is a net cooling/heating function.

The mass fraction of cold gas within the gas cell, $x\equiv\rho_{\rm c}/\rho_{\rm g}$, is computed as   
\begin{equation}
x=1+\frac{1}{2y}-\sqrt{\frac{1}{y}+\frac{1}{4y^2}},
\end{equation}
where 
\begin{equation}
y\equiv\frac{t_{\rm SF}\Lambda_{\rm net}(\rho_{\rm g},u_{\rm h})}{\rho_{\rm g}\left[\beta u_{\rm SN}-\left(1-\beta\right)u_{\rm c}\right]}.
\end{equation}
The quantities of $x$ and $y$ depend only on the total density $\rho_{\rm g}$ of the cell provided that $\Lambda_{\rm net}$ is a function of $\rho_{\rm g}$ and $u_{\rm h}$. In a practical range of $\rho_{\rm g}$ in galaxy simulations, $x$ is not significantly lower than unity, therefore splitting a gas cell into the two phases does not largely decreases an amount of gas allowed to form stars.

The effective pressure of the gas cell given by equation (\ref{p_eff}) is written as 
\begin{equation}
P_{\rm eff}=(1-\gamma)\rho_{\rm g}\left[xu_{\rm c}+\left(1-x\right)u_{\rm h}\right],
\label{p_eff2}
\end{equation}
and it is a function only of $\rho_{\rm g}$. The thermal energy is given as $u_{\rm eff}=\left[xu_{\rm c}+\left(1-x\right)u_{\rm h}\right]$. In our simulations, the two-phase ISM model thus enforces a temperature floor and a pressure floor according to the barotropic EOS (equation \ref{p_eff2}) on dense gas with $\rho_{\rm g}>\rho_{\rm EOS}$. Because the SN feedback of unresolved stars is included, the effective temperature $u_{\rm eff}$ monotonically increases with $\rho_{\rm g}$. This means that the effective EOS of this model is harder than an isothermal EOS. Hence, adopting this model can stabilise dense gas and simultaneously prevents unphysical fragmentation that can occur at small scales of resolution limits (see Section \ref{singlephasemodel}). The above ISM model we adopt is essentially the same as that presented in \citet[][]{sh:03} except the assumed parameters and IMF.

The main purpose of the two-phase ISM model is to take into account unresolved star formation and SN feedback in sub-resolution scales. Hence, we consider that $\rho_{\rm EOS}$ should be set equal to the star-formation criterion in the simulations: $\rho_{\rm EOS}=n_{\rm H,th}=0.13~{\rm cm^{-3}}$.\footnote{However, it is not necessarily required to set $\rho_{\rm EOS}=\rho_{\rm th}$. We also run simulations with a higher $\rho_{\rm th}$ while $\rho_{\rm EOS}$ is fixed (see Section \ref{potentialproblem}). } Left panels of Fig. \ref{PhaseDiagrams} show phase diagrams in our simulation using this model. In the diagrams, the dense gas is almost smoothly connected to low-density gas at the boundary $\rho_{\rm g}=\rho_{\rm EOS}$. The two-phase ISM model has been implemented in the original version of {\sc Arepo} and used in Illustris \citep{Illustris} and Illustris-TNG \citep{TNG} simulations, as well as in Auriga simulations \citep{ggm:17} which are zoom-in simulations using the same code. 

\subsubsection{Polytrope ISM model}
\label{polytropemodel}
For smoothed particle hydrodynamics (SPH) simulations, \citet{sd:08} have proposed an ISM model in which a polytropic EOS is imposed on dense gas, i.e. $P_{\rm eff}\propto\rho_{\rm g}^n$, with the polytrope index $n=4/3$. This choice is motivated because a Jeans mass is independent of $\rho_{\rm g}$ when $n=4/3$ while a Jeans length becomes shorter with increasing $\rho_{\rm g}$. Hence, it is expected that the index $n=4/3$ allows a gas cloud to collapse but prevents it to fragment into smaller clouds. Because clouds in simulations can artificially fragment when a mass of a gas element is larger than a Jeans mass \citep{bb:97}, it is desirable for simulations to make a Jeans mass invariant with $\rho_{\rm g}$. Their ISM model with $n=4/3$ has been used in not only SPH simulations but also adaptive mesh refinement (AMR) simulations \citep{d:14,c:18}. We refer to this model as `polytrope ISM model', examine it in our simulations using the moving-mesh code.

The polytrope ISM model enforces a pressure floor $P_{\rm eff}\propto\rho_{\rm gas}^{4/3}$ on dense gas with $\rho_{\rm g}>\rho_{\rm EOS}$, and its temperature is also adjusted to $u_{\rm eff}=P_{\rm eff}/[\rho_{\rm g}(1-\gamma)]$. The normalisation is given as $T_{\rm eff}=8\times10^3~{\rm K}$ at $\rho_{\rm g}=0.1~{\rm cm^{-3}}$ \citep[e.g.][]{s:15}. As is the case of the two-phase model, we set $\rho_{\rm EOS}=\rho_{\rm th}$. The dense gas is stabilised by the effective pressure and prevented from artificial fragmentation in small scales resolved scarcely. In this model, we assume that all mass within a cell with $\rho_{\rm g}>\rho_{\rm th}$ can form stars therefore set $x=1$ in computing an SFR of the cell.

Phase diagrams of the polytrope ISM model are shown in Fig. \ref{PhaseDiagrams} (centre). Although the gas distribution is similar to that in the two-phase-ISM run, the effective EOS is softer (shallower slopes) especially in the range of $n_{\rm H,th}<n_{\rm H}\lsim1~{\rm cm^{-3}}$. The polytrope model has been used in the cosmological simulations of OWLS \citep{owls} and EAGLE projects \citep{s:15}. In addition, their zoom-in simulations of the APOSTLE project \citep{apostle} --- which focuses on analogues of the Milky Way and Andromeda galaxies \citep{s:16} --- are also performed using the same code\footnote{OWLS, EAGLE and APOSTLE simulations were performed with an $N$-body Tree-PM, SPH code {\sc GADGET3} \citep[last described in][]{gadget} modified by the projects.} of the EAGLE simulations. Horizon-AGN simulation also uses this ISM model \citep{d:14,c:18}.

\subsubsection{Single-phase ISM model}
\label{singlephasemodel}
A straightforward method to compute hydrodynamics of dense gas is not to use any ISM models. In other words, gas is not distinguished between $\rho_{\rm g}>\rho_{\rm EOS}$ and $<\rho_{\rm EOS}$ in hydrodynamics. In this study, we refer to this method as `single-phase ISM model', in contrast with the two-phase model described in Section \ref{twophasemodel}. In this model, gas in our simulations can cool according to the cooling/heating function until $5~{\rm K}$ regardless of its density. 

Unlike the two-phase and the polytrope models, the absence of pressure support for dense gas can, however, lead cold gas to fragment artificially in small scales at resolution limits. To avoid this, we need to impose a floor pressure. Using AMR simulations for a gas cloud in isolation, \citet{tkm:97} have demonstrated that a Jeans length should be resolved with at least 4 cells, otherwise a gas cloud fragments artificially: the Truelove condition. In more realistic situations for galaxy formation, \citet{cdb:10} have performed AMR cosmological simulations for clumpy galaxies and shown that the number of clumps in a disc galaxy converges when a pressure floor is imposed so that a Jeans length $\lambda_{\rm J}$ is always resolved with at least 7 cells. This means $N_{\rm cell}\equiv\lambda_{\rm J}/\Delta=7$, where $\Delta$ is a spatial resolution limit (the minimum cell size in an AMR simulation). The Jeans length is estimated as $\lambda_{\rm J}^2=\pi c_{\rm snd}^2/(G\rho_{\rm g})$, where $G$ is the gravitational constant, and sound speed is given as $c_{\rm snd}^2=\gamma P/\rho_{\rm g}$. Accordingly, the floor pressure to stabilise physical scales smaller than $\lambda_{\rm J}$ is given as 
\begin{equation}
P_{\rm eff}=\frac{G\rho_{\rm g}^2N_{\rm cell}^2\Delta^2}{\pi\gamma}.
\label{TLC}
\end{equation}
Following \citet{cdb:10}, we adopt $N_{\rm cell}=7$ in our simulations since it is more stringent than the Truelove condition. We assume $\Delta=0.3~{\rm kpc}$ (comoving). Note, however, that it is unclear what value of $\Delta$ is appropriate in moving-mesh simulations since basically there is no minimum limit of a cell size in our moving-mesh scheme. The value of $\Delta=0.3~{\rm kpc}$ is comparable to diameters of the smallest cells in our simulations with the two-phase ISM models at $z\sim2$--$1$; this cell size corresponds to a density of $n_{\rm H}=\tilde{m}_{\rm g}/[4\pi(\Delta/2)^3/3]\sim10~{\rm cm^{-3}}$. Thus, the value of $\Delta=0.3~{\rm kpc}$ is not too small for the resolution of our simulations. We confirm, in addition, that results of our simulations using this model hardly change even if we adopt $\Delta=1.5~{\rm kpc}$ and even if $\Delta$ is set equal to an actual diameter of each cell. We do not impose temperature floors on gas cells.\footnote{This treatment is the same as in VELA simulations \citep{cdb:10,cdm:11,ckk:14}.} The pressure floors are imposed on all gas cells irrespective of their densities.\footnote{Gas with $\rho_{\rm g}<\rho_{\rm th}$ rarely reach a pressure lower than the floor.} In this model, all mass within a cell with $\rho_{\rm g}>\rho_{\rm th}$ is able to form stars; therefore we set $x=1$ for computing an SFR of a cell in our simulations.

\begin{figure*}
  \includegraphics[bb=0 0 2540 1129, width=\hsize]{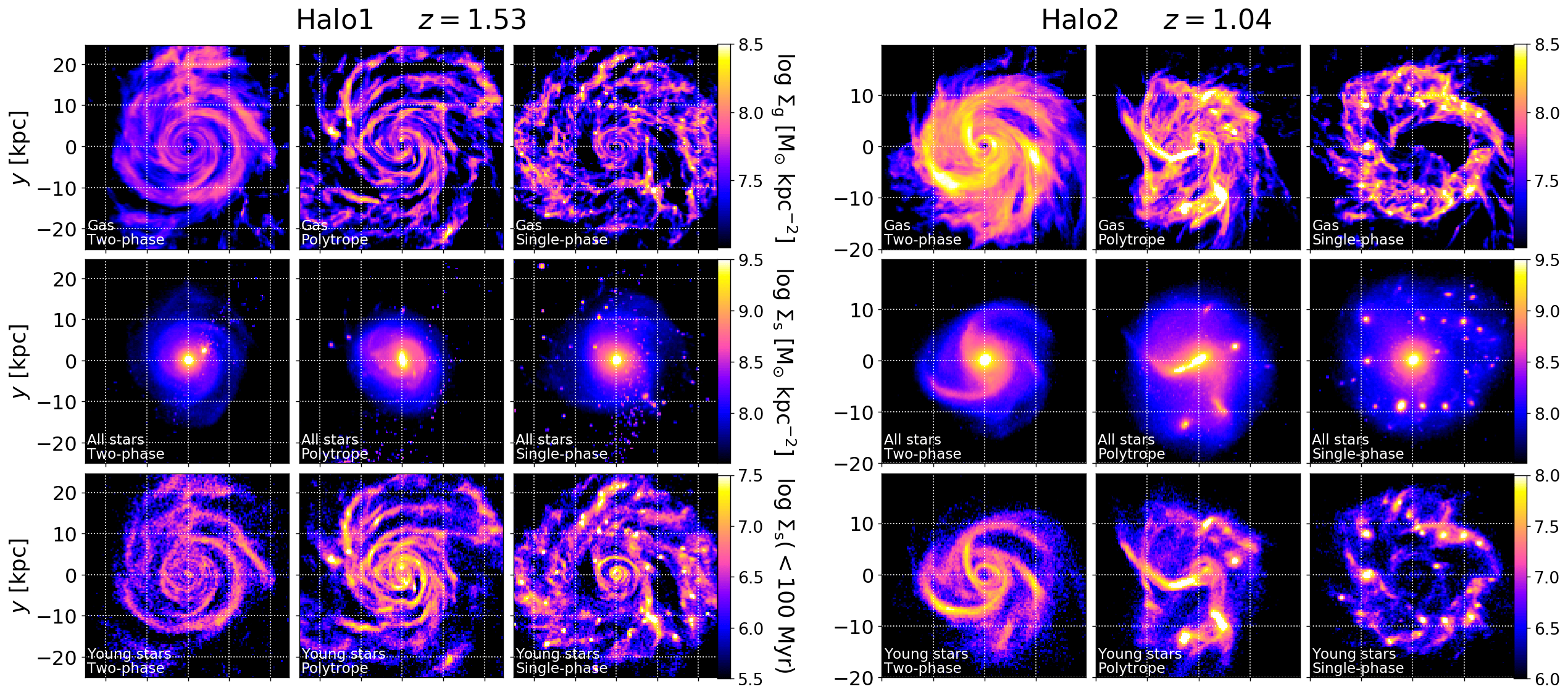}
  \caption{Face-on maps of surface densities of gas (top), all stars (middle) and stars younger than $100~{\rm Myr}$ (bottom) in our simulations with the two-phase (left), polytrope (centre) and single-phase (right) ISM models. Left and right sets of panels show snapshots of Halo 1 and 2 at $z=1.53$ and $1.04$, respectively.}
  \label{snapshots1}
\end{figure*}

\begin{figure*}
  \includegraphics[bb=0 0 3682 1129, width=\hsize]{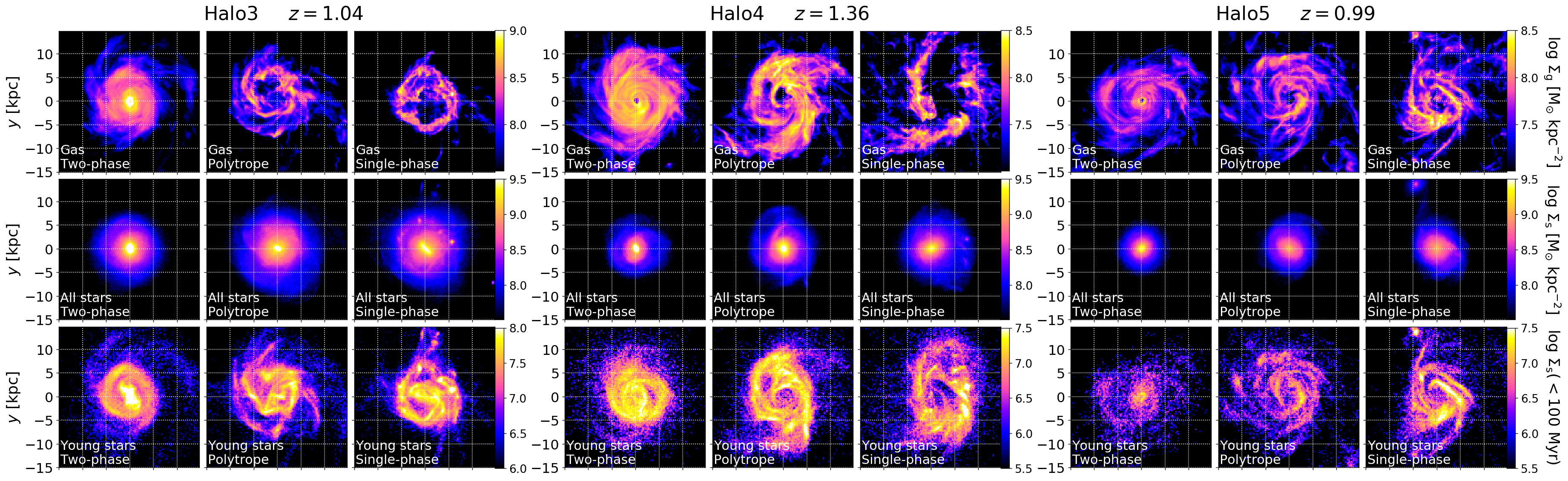}
  \caption{Same as Fig. \ref{snapshots1} but for Halo 3 (left), 4 (centre) and 5 (right) at $z=1.04$, $1.36$ and $0.99$, respectively.}
  \label{snapshots2}
\end{figure*}

Phase diagrams in our simulation using the single-phase model with $\Delta=0.3~{\rm kpc}$ is shown in Fig. \ref{PhaseDiagrams} (right). Most of star-forming gas with $\rho_{\rm g}\gsim\rho_{\rm th}$ distributes along two sequences. A substantial amount of the dense gas evolves nearly isothermally with $T\simeq10^4~{\rm K}$, i.e. $P\propto\rho_{\rm g}$. On the other hand, cold dense gas follows the sequence imposed by equation (\ref{TLC}), i.e. $P\propto\rho_{\rm gas}^2$ in the bottom panel. The isothermal sequence corresponds to an EOS softer than the other two models. The presence of the isothermal sequence at $T\simeq10^4~{\rm K}$ is consistent with results of previous studies in which their simulations assume similar single-phase ISM models \citep[e.g.][]{ssk:06}. The single-phase ISM model, i.e. not enforcing a modelled EOS on dense gas, is used in various simulations, e.g. FIRE/FIRE-2 \citep{FIRE,FIRE2}, NIHAO \citep{NIHAO} and VELA \citep{cdb:10,cdm:11,ckk:14} simulations.

\section{Results}
\label{result}
\subsection{Influence on galactic clumpiness}

\begin{table*}
  \caption{Physical properties of the simulated galaxies in the snapshots shown in Figs. \ref{snapshots1} and \ref{snapshots2}. The left three columns are halo IDs and redshifts of the snapshots, the total masses ($M_{200}$) and sizes ($R_{200}$) of the systems. The values of $M_{200}$ and $R_{200}$ are measured with the friend-of-friend grouping algorithm and almost independent of the ISM models. The right six columns are for the runs with the ISM models, the total stellar masses ($M_{\rm s}$), gas masses ($M_{\rm g}$), SFR, the mean stellar metallicities ($Z_{\rm s}$) and the numbers of clumps whose total baryon masses are $>10^8~{\rm M_\odot}$ ($N_{\rm clump}$). We assume the solar metallicity $Z_\odot=0.0127$. These values are measured within regions inside $0.15R_{200}$ for each run, except $N_{\rm clump}$ that are counted within the regions shown in Figs \ref{snapshots1} and \ref{snapshots2}.}
  \label{paramlist}
  \begin{tabular}{cccccccccc}
    \hline
     run & $M_{200}~[{\rm M_\odot}]$ &  $R_{200}~[{\rm kpc}]$ & ISM model & $M_{\rm s}~[{\rm M_\odot}]$ & $M_{\rm g}~[{\rm M_\odot}]$ & SFR $[{\rm M_\odot~yr^{-1}}]$ & $Z_{\rm s}~[{\rm Z_\odot}]$ & $N_{\rm clump}$ &\\
    \hline
    \multirow{3}{*}{\shortstack{Halo1\\($z=1.53$)}} &  & & two-phase & $1.69\times10^{11}$ & $1.08\times10^{11}$ & $48.4$ & $1.66$ & 0 &\\
    & $5.16\times10^{12}$ & $205$ & polytrope & $1.98\times10^{11}$ & $1.05\times10^{11}$ & $91.8$ & $1.95$ & 8 &\\
    &  & & single-phase & $2.14\times10^{11}$ & $1.09\times10^{11}$ & $102$ & $2.10$ & 49 &\\
    \hline
    \multirow{3}{*}{\shortstack{Halo2\\($z=1.04$)}} & & & two-phase& $1.82\times10^{11}$ & $1.27\times10^{11}$ & $86.6$ & $1.73$& 0 &\\
    &  $5.47\times10^{12}$ & $250$ & polytrope & $2.63\times10^{11}$ & $1.18\times10^{11}$ & $175$ & $2.20$ & 6 &\\
    & & & single-phase & $2.32\times10^{11}$ & $1.07\times10^{11}$ & $101$ & $2.08$ & 35 &\\
    \hline
    \multirow{3}{*}{\shortstack{Halo3\\($z=1.04$)}} & & & two-phase& $8.49\times10^{10}$ & $1.06\times10^{11}$ & $76.0$ & $1.47$ & 0 &\\
    & $4.46\times10^{12}$ & $234$ & polytrope & $1.10\times10^{11}$ & $8.22\times10^{10}$ & $55.8$ & $1.66$ & 7 &\\
    & & & single-phase & $1.14\times10^{11}$ & $7.02\times10^{10}$ & $62.1$ & $1.64$ & 9 &\\
    \hline
    \multirow{3}{*}{\shortstack{Halo4\\($z=1.36$)}} & & & two-phase & $4.08\times10^{10}$ & $4.51\times10^{10}$ & $26.4$ & $1.28$ & 0 &\\
    & $1.85\times10^{12}$ & $154$ & polytrope & $5.70\times10^{10}$ & $4.81\times10^{10}$ & $35.0$ & $1.49$ & 0 &\\
    & & & single-phase & $5.26\times10^{10}$ & $3.99\times10^{10}$ & $18.3$ & $1.36$ & 0 &\\
    \hline
    \multirow{3}{*}{\shortstack{Halo5\\($z=0.99$)}} & & & two-phase & $3.06\times10^{10}$ & $3.12\times10^{10}$ & $8.32$ & $1.31$ & 0 &\\
    & $1.86\times10^{12}$ & $178$ & polytrope & $3.98\times10^{10}$ & $3.74\times10^{10}$ & $13.4$ & $1.35$ &  0 &\\
    & & & single-phase & $4.69\times10^{10}$ & $4.25\times10^{10}$ & $18.5$ & $1.41$ & 2 &\\
    \hline
  \end{tabular}
\end{table*}

\begin{figure}
  \includegraphics[bb=0 0 1304 2098, width=\hsize]{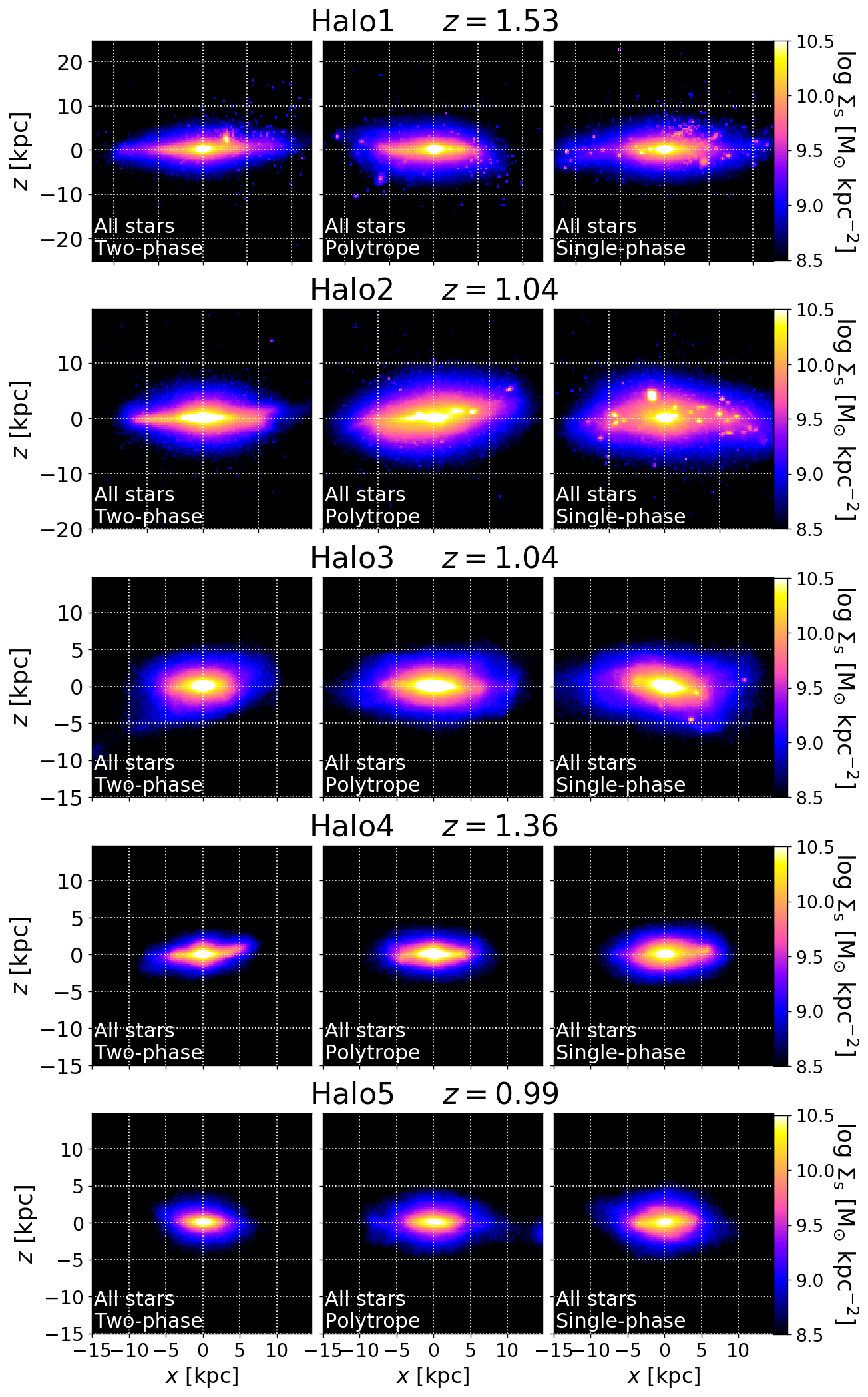}
  \caption{Edge-on maps of surface densities of all stars in the snapshots shown in Figs. \ref{snapshots1} and \ref{snapshots2}. From left to right, the panels show the runs with the two-phase, polytrope and single-phase ISM models, respectively.}
  \label{EdgeOn}
\end{figure}
Figs. \ref{snapshots1} and \ref{snapshots2} show face-on surface density maps of gas, all stars and stars younger than $100~{\rm Myr}$ for the five haloes at redshifts $z\simeq1$--$1.5$ with  three ISM models. At these output times, we find no significant impact of merger events that can strongly disturb the disc structures. Table \ref{paramlist} lists properties of the galaxies. Fig. \ref{EdgeOn} shows edge-on surface density maps of all stars for the same snapshots. Their vertical scale heights are significantly smaller than the disc scale lengths. Because the face-on stellar distributions are nearly circular in Figs. \ref{snapshots1} and \ref{snapshots2}, the simulated galaxies appear to be discs. We discuss the stellar disc structures in Section \ref{ang}.

Figs. \ref{snapshots1} and \ref{snapshots2} demonstrate that the disc structures are clearly affected by the ISM models. In the top panels, the gas density distributions become highly clumpy in the cases of the single-phase and the polytrope ISM models. In these runs of Halo 1 and 2, we find that the discs are in clumpy states in $z\lsim1.5$--$2$. On the other hand, no massive clumps form in the two-phase-ISM runs, and spiral arms appear less prominent than in the runs with the other ISM models. Thus, the ISM models gas can remarkably change the clumpiness of galactic discs at $z\sim1$-$1.5$. Especially, the difference is clearer in massive galaxies: Halo 1 and 2. The impact of switching the ISM models may be small in less massive galaxies: Halo 3, 4 and 5. The single-phase ISM model appears to produce more clumps than the polytrope model. The giant clumps have high SFRs, therefore can be seen more prominently in density maps of young stars (bottom panels), which can be taken as proxy for star-forming lights such as H$\alpha$ and UV emission.\footnote{H$\alpha$ emission line is sensitive only to stars younger than $\sim10~{\rm Myr}$, and rest-frame UV light traces stars whose ages are $\sim10$--$100~{\rm Myr}$. In this sense, the stars younger than $100~{\rm Myr}$ would be consistent the best with UV observations; note that we lack dust obscuration.} Moreover, these clumps are massive enough to be gravitationally bound structures, and seen in density maps of all stars (middle panels). We find that they survive over periods of their orbital time-scales in the simulations. Note, however, that lifetimes of clumps would depend on strength of stellar feedback as we mention in Section \ref{Intro}. Our results show that formation efficiency of giant clumps in galactic discs can significantly depend on EOS of dense gas.

\begin{figure}
  \includegraphics[bb=0 0 1170 786, width=\hsize]{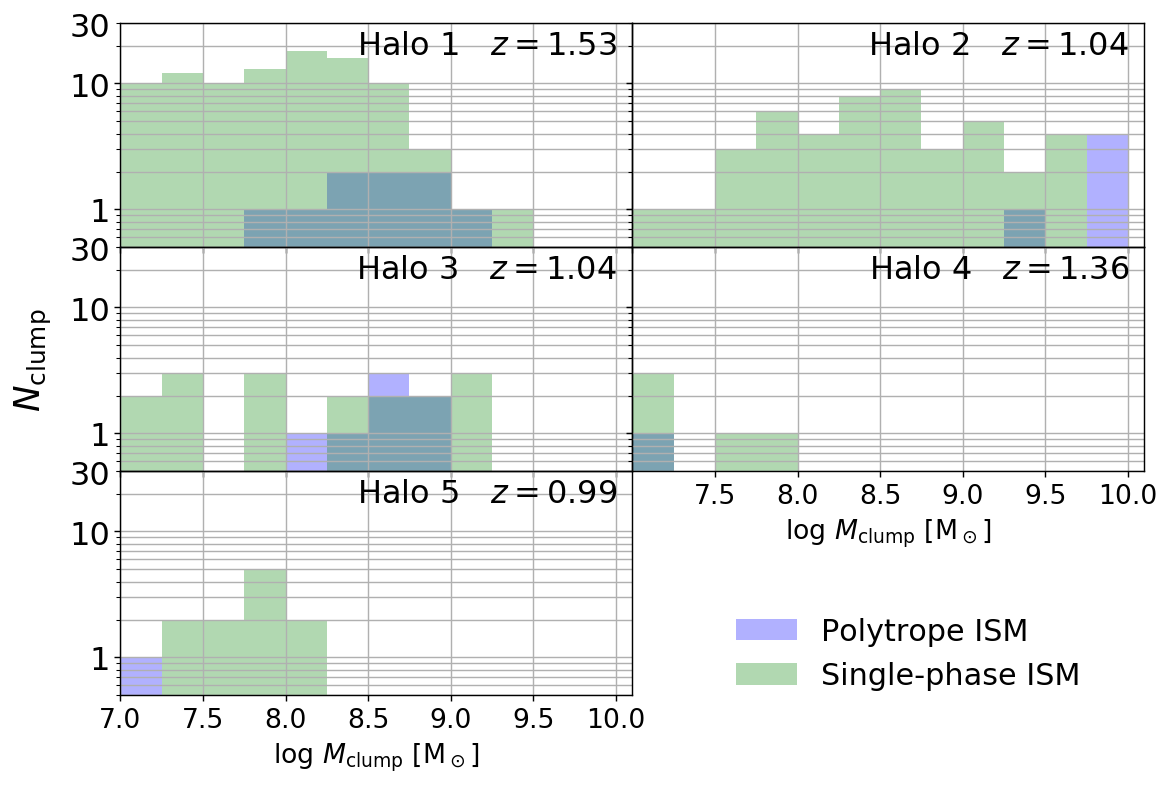}
  \caption{Mass functions of clumps identified in three-dimensional gas distributions in the snapshots shown in Figs. \ref{snapshots1} and \ref{snapshots2}. The value of $M_{\rm clump}$ is defined as the total baryon mass of each clump after removing unbound matters. We exclude ex-situ clumps whose DM fractions are higher than $0.1$. There are no clumps more massive than $10^{5.5}~{\rm M_\odot}$ in the snapshots of the two-phase-ISM runs. The numbers of clumps $>10^8~{\rm M_\odot}$ are listed in the rightmost column of Table \ref{paramlist}.}
  \label{Mclump}
\end{figure}
To the snapshots shown in Figs. \ref{snapshots1} and \ref{snapshots2}, we apply a clump identification scheme that is similar to those used in \citet{mdc:13,mdc:17} and \citet{idm:16} (see Appendix \ref{clumpdetection} for the details). We compute the total baryon masses $M_{\rm clump}$ of clumps identified in the regions depicted in Figs. \ref{snapshots1} and \ref{snapshots2} within the vertical heights of $\pm5~{\rm kpc}$ from the disc planes. We then exclude clumps whose DM mass fractions are greater than $0.1$ since such clumps are considered to be `ex-situ' clumps, i.e. satellite galaxies. Histograms of $M_{\rm clump}$ are illustrated in Fig. \ref{Mclump}, and the numbers of the clumps with $M_{\rm clump}>10^8~{\rm M_\odot}$ are listed in the rightmost column of Table \ref{paramlist}. As expected in Figs. \ref{snapshots1} and \ref{snapshots2}, the runs with the single-phase ISM model form larger numbers of clumps than those with the other models. On the other hand, no clumps with $M_{\rm clump}>10^{5.5}~{\rm M_\odot}$ are identified in the snapshots of the two-phase-ISM runs in all haloes. Thus, using the single-phase and the polytrope ISM models significantly increases the numbers of clumps in our cosmological simulations. This result is explained by the difference of EOS in the models. Namely, since a softer EOS is naturally expected to allow small-scale structures to grow due to relatively weaker pressure response against density variation \citep[e.g.][]{sd:08,hsb:13,msv:19}.

\subsection{Influence on global properties of galaxies}
 In spite of the clear differences in clumpiness, especially in Halo 1 and 2, we find that switching the ISM models hardly changes global properties of the simulated galaxies. In Table \ref{paramlist}, the total stellar and gas masses, SFRs and metallicities of the galaxies in the snapshots are almost the same between the runs with the three ISM models in all haloes.

\begin{figure}
  \includegraphics[bb=0 0 1177 1228, width=\hsize]{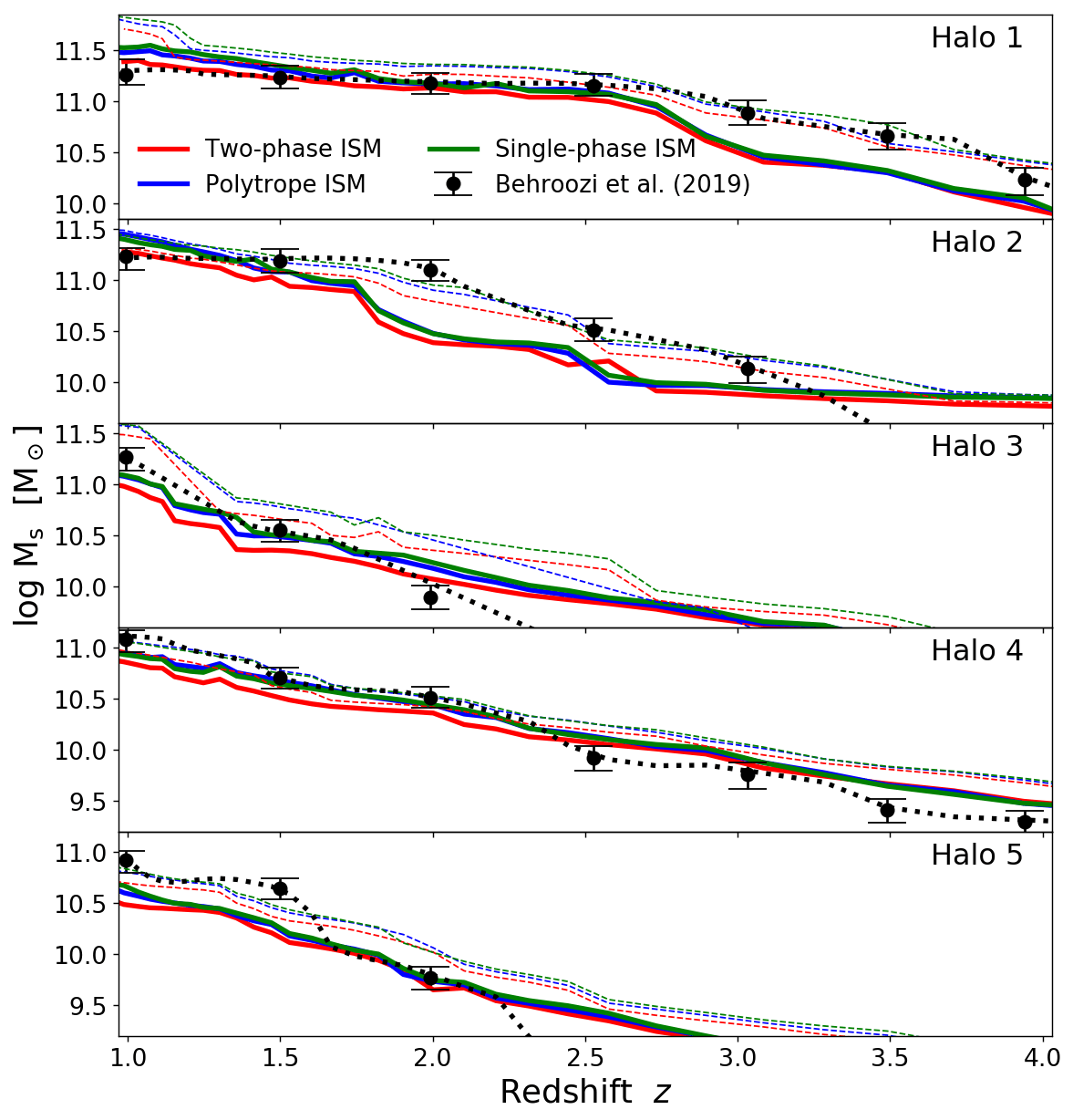}
  \caption{Redshift-evolution of the total stellar masses of Halo 1 to 5. Thick solid and thin dotted lines show the stellar masses enclosed within radii of $0.15R_{200}$ and $R_{200}$, respectively. Black filled circles with error bars indicate stellar masses estimated from the stellar-to-halo mass relations \citep{bwh:19} adopted to $M_{200}$ in our simulations. Black dotted lines are drawn from their fitting functions for the stellar mass-halo mass relations.}
  \label{Ms}
\end{figure}

\begin{figure}
  \includegraphics[bb=0 0 1183 1235, width=\hsize]{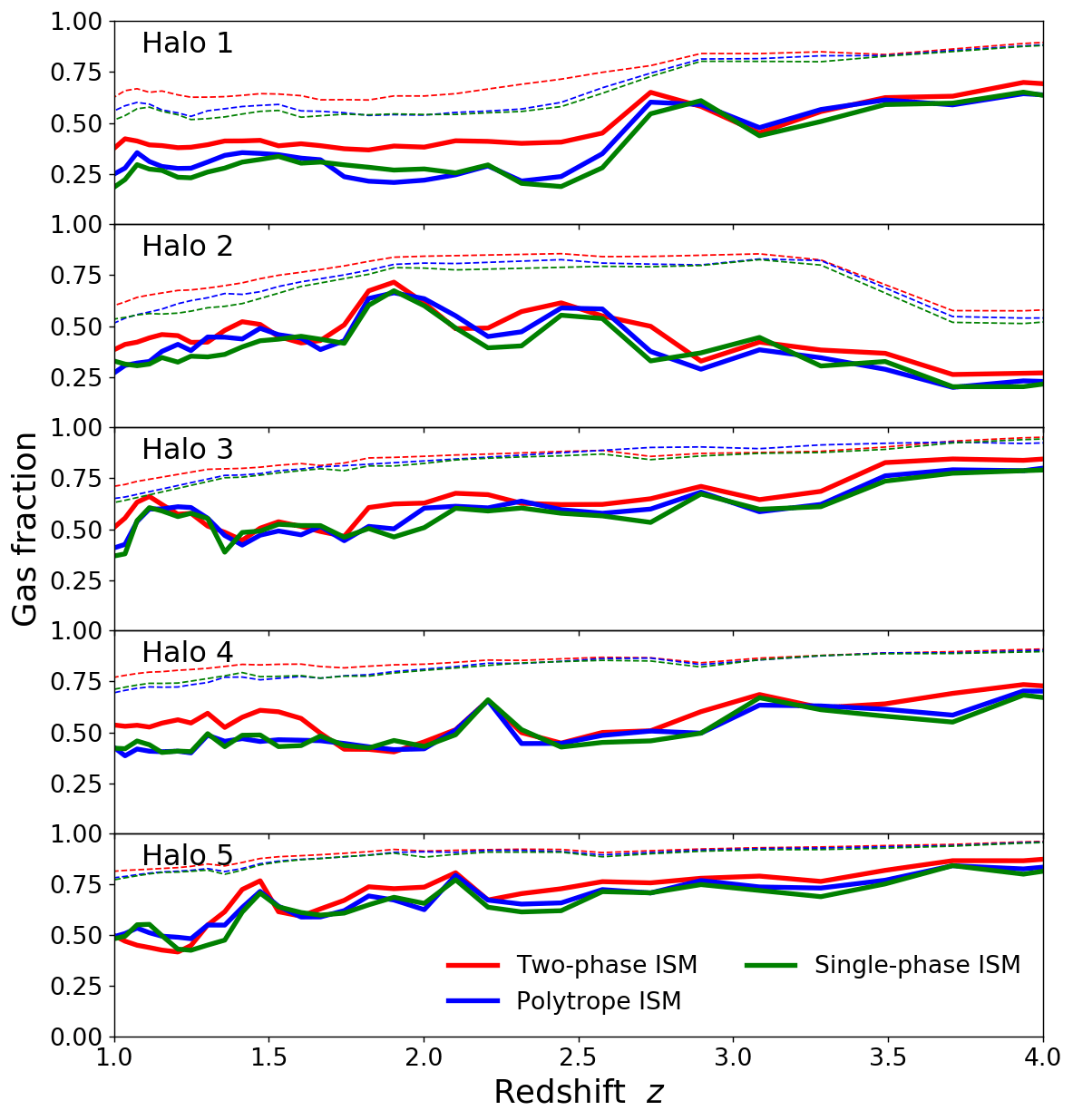}
  \caption{Redshift-evolution of gas fractions, $M_{\rm g}/(M_{\rm s}+M_{\rm g})$, of Halo 1 to 5. Thick solid and thin dotted lines indicate values measured within radii of $0.15R_{200}$ and $R_{200}$, respectively.}
  \label{fgas}
\end{figure}

\begin{figure}
  \includegraphics[bb=0 0 1175 1228, width=\hsize]{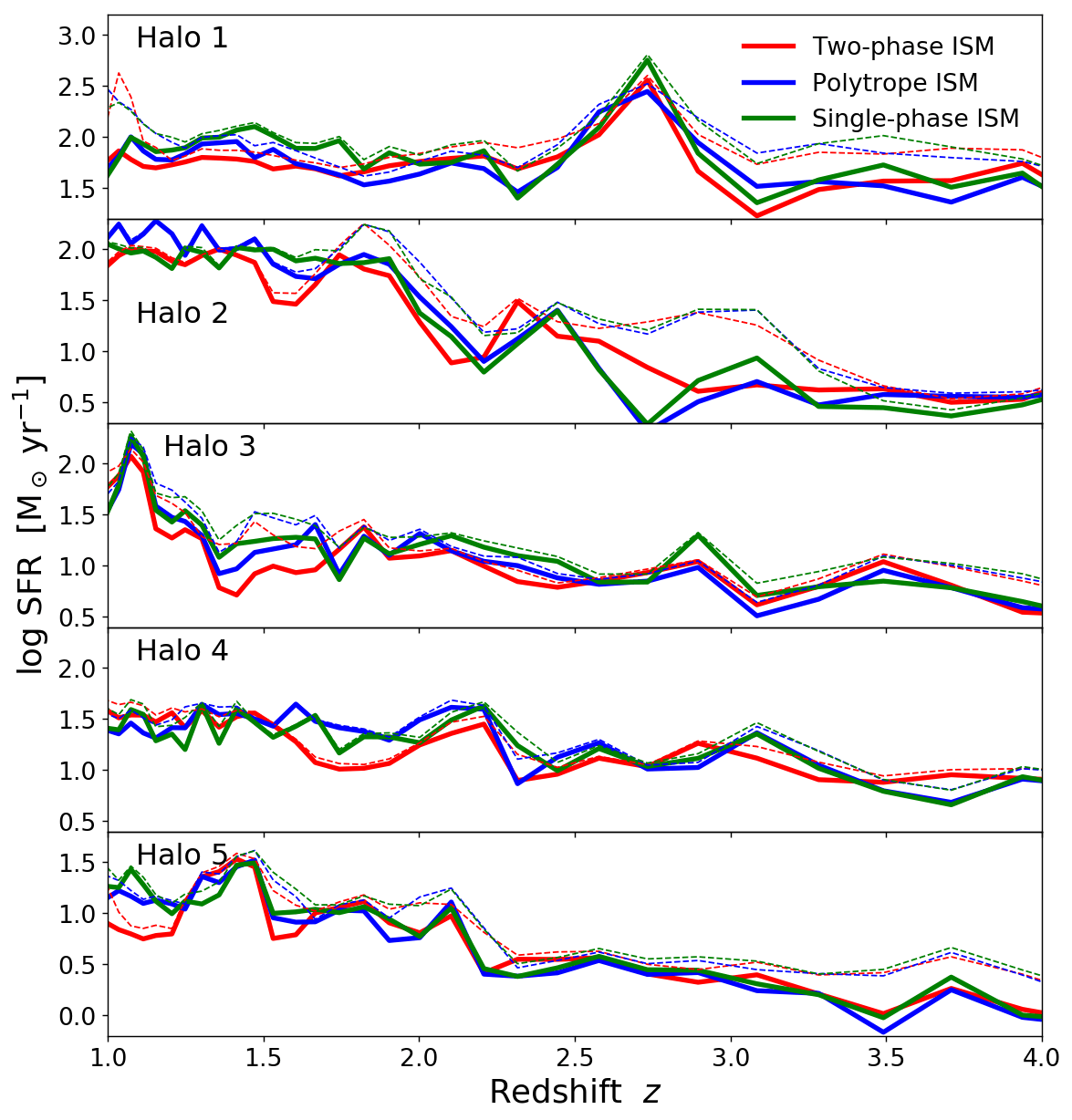}
  \caption{Redshift-evolution of the total SFRs of Halo 1 to 5. Thick solid and thin dotted lines indicate values measured within radii of $0.15R_{200}$ and $R_{200}$, respectively.}
  \label{sfr}
\end{figure}

Figs \ref{Ms}, \ref{fgas} and \ref{sfr} show the time-evolution of the total stellar masses, gas fractions and SFRs in our simulations. The solid lines indicate these values measured within regions of $r<0.15R_{200}$ which enclose most of baryon belonging to the central galaxies while excluding most of satellite galaxies. In Fig \ref{Ms}, the stellar masses do not appear to differ significantly. We compare $M_{\rm s}(<0.15R_{200})$ between the three ISM models for all the five haloes from $z=4$ to $1$ and find the largest difference to be only 63 per cent. It should be noted, however, that in Fig. \ref{Ms} the stellar masses in the two-phase-ISM runs seem to be systematically lower than those in the other runs. For comparison, in Fig. \ref{Ms}, we also plot stellar masses estimated from results of the UniverseMachine \citep[][]{bwh:19}, which is based on an abundance-matching technique combining an observed stellar mass function and a DM mass function obtained from a cosmological $N$-body simulation.\footnote{The UniverseMachine provides stellar mass-halo mass relations for various subgroups: star-forming/quenched and central/satellite galaxies. In addition, it keeps track of two stellar masses: the `true' and `observed' masses. The true mass is computed by integrating past SFRs subtracted with stellar mass loss. The observed mass is determined from observations. In Fig. \ref{Ms}, we use their results of true stellar masses for star-forming galaxies without distinguishing central and satellite galaxies.}  In the figure, the stellar masses are estimated by multiplying $M_{200}$ by the stellar-to-halo mass ratios obtained from the UniverseMachine. The stellar masses in our simulations are mostly consistent with the estimations from the UniverseMachine. However, in $z\gsim3$ (Halo 1) and $z\simeq2$--$3$ (Halo 2), the UniverseMachine predicts stellar masses significantly larger than those within $0.15R_{200}$ in our simulations. In these redshift ranges of the runs, $M_{\rm s}(<0.15R_{200})$ are discrepant from $M_{\rm s}(<R_{200})$ in our simulations, and the UniverseMachine is rather consistent with $M_{\rm s}(<R_{200})$. The large difference between $M_{\rm s}(<0.15R_{200})$ and $M_{\rm s}(<R_{200})$ implies that there could be other massive galaxy/galaxies within the common halo, and our analysis scheme may not trace the most massive one.\footnote{We define the galactic centre as the DM particle that resides at the lowest potential.} In addition, there may be over-production of stars in $z\gsim2.5$ in Halo 3 and 5, and under-production in $z\lsim1.5$ in Halo 5. However, the discrepancy from the UniverseMachine is seen in all runs with the three ISM models. Namely, it is not caused by changing the ISM models.

Amount of gas is thought to play a key role in clump formation since clumpy galaxies are generally observed to be highly gas-rich \citep[e.g.][]{g:08,gnj:11,w:12,w:12b,fgb:14}. Fig \ref{fgas} does not show significant differences in gas fractions between the three ISM models. In Halo 1 at $z\lsim2.5$ and Halo 4 at $z\lsim1.5$, the two-phase-ISM runs appear to indicate systematically higher gas fractions than the runs with the other ISM models. This may reflect the lower stellar masses shown in Fig \ref{Ms}. It is worthy to mention that the clumpy galaxies in the runs of Halo 1 and 2 with the polytrope and the single-component ISM models have lower gas fractions than the non-clumpy discs in the two-phase-ISM runs. This means that the absence of giant clumps in the two-phase-ISM runs is not attributed to poverty of gas. Our simulated galaxies have gas fractions of $f_{\rm g}\simeq0.2$--$0.4$ in their clumpy phases, which are significantly higher than the typical values of local spiral galaxies, $f_{\rm g}\simeq0.1$. However, Halo 3, 4 and 5 are not (or less) clumpy even in the case of the single-phase ISM model although they have even higher gas fractions $f_{\rm g}\simeq0.5$. This implies that a high gas fraction is not the sufficient condition of clump formation, may be one of necessary conditions. The two-phase-ISM runs may have slightly but systematically higher gas fractions than the other runs; this trend is clear in Halo 1 ($z\lsim2.5$) and Halo 4 ($z\lsim1.5$). This slightly higher $f_{\rm g}$ would be related to the systematically lower $M_{\rm s}$ of the two-phase-ISM runs (Fig \ref{Ms}). The systematic trend would be attributed to the cold gas fraction $x$ and the effective pressure $P_{\rm eff}$. The two-phase model allows $x$ to be slightly lower than unity and gives the highest pressure floor among the three models (Fig. \ref{PhaseDiagrams}).

Formation of giant clumps may significantly increase SFRs in galaxies. \citet{ggf:11} have observationally estimated the total SFR within giant clumps in a galaxy at $z\sim2$ to account for nearly fifty per cent of that within the entire galaxy.\footnote{Note that estimations of clump sizes and masses can significantly depend on observational resolutions \cite[e.g.][]{dsc:17,csr:18}. This would also be the case for estimations of SFRs of clumps.} Fig \ref{sfr} shows dependence of SFRs on the ISM models. The total SFRs fluctuate by a factor of $2$--$3$ in short periods in our simulations. However, we find no systematic difference in the SFRs between the three ISM models, i.e. the difference between the runs are smaller than the short-period fluctuation. In addition, enhancement of the total SFRs is not clearly seen in the clumpy phases in Halo 1 and 2 at $z\lsim1.5$--$2$. This implies that clump formation itself would not significantly increase the total SFR in a galaxy by a factor of $\gsim2$--$3$. It has been observed that $\sim50$ per cent of star-forming galaxies at $z\simeq2$--$3$ have clumpy morphologies \citep{tkt:13II,mkt:14,gfb:14,sok:16,bmo:17}. We expect that the high SFRs of the observed clumpy galaxies would not be driven by their clump formation.

\subsection{Influence on disc properties and relationships}
\subsubsection{Angular momentum of a stellar disc}
\label{ang}
Early simulations suffered from the over-cooling problem and the angular-momentum catastrophe. They resulted in too compact discs with low angular momenta \citep[e.g.][]{katz:91}. Although modern simulations have overcome this problem by increasing resolutions and improving recipes of stellar feedback such as SNe, changing the ISM models may result in recurrence of this problem.

\begin{figure}
  \includegraphics[bb=0 0 713 488, width=\hsize]{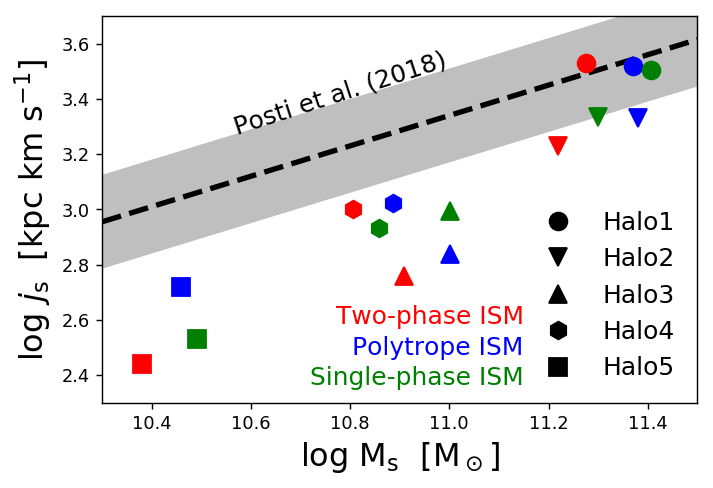}
  \caption{Relationship between the total masses and specific angular momenta of stellar components in our simulations at the redshift $z=1$. Symbols with different colours correspond to the runs with the three ISM models. The dashed line delineates the observed relation of \citet{pfd:18} and \citet{mfp:19}, and the shaded region indicates the $1\sigma$ orthogonal intrinsic scatter of their sample.}
  \label{momenta}
\end{figure}

We define a specific angular momentum of a stellar disc as 
\begin{equation}
j_{\rm s}(R)=\sum_{R_i<R}m_{{\rm s},i}R_iv_{\phi,i}\bigg/\sum_{R_i<R}m_{{\rm s},i},
\label{AngMom}
\end{equation}
where $m_{{\rm s},i}$, $v_{\phi,i}$ and $R_i$ are mass, azimuthal velocity and projected distance from the galactic centre of $i$-th stellar particle, respectively. We exclude stellar particles above and below the regions within $5~{\rm kpc}$ from the disc plane.\footnote{We confirm that $j_{\rm s}$ hardly changes even if the vertical cutoff is set to $|z|=10~{\rm kpc}$ or $|z|=R$.} The values of $j_{\rm s}$ usually increase with $R$ but converge at a large $R$. Since we find that $R=0.15R_{200}$ is large enough for the convergence, we compute $j_{\rm s}$ within this radius.\footnote{In the single-phase-ISM run of Halo 5 at $z=1$, there is a infalling satellite at $R\simeq15~{\rm kpc}$ (see Fig. \ref{snapshots2}). To avoid this satellite, we compute $j_{\rm s}$ at $R=10~{\rm kpc}$ for all runs of Halo 5. We confirm that this radius is large enough for the convergence in these runs.}

We compute $j_{\rm s}$ of the simulated stellar discs at $z=1$. Fig. \ref{momenta} shows that $j_{\rm s}$ does not largely depend on the ISM models although it differs by a factor of $\sim2$ in the runs of Halo 3 and 5. The clumpy discs in Halo 1 and 2 with the polytrope and the single-phase models do not indicate systematically lower or higher angular momenta. Thus, although using the ISM models with softer EOS can result in high clumpiness, it does not catastrophically decrease the angular momenta of the stellar discs. 

Correlation between stellar masses and angular momenta of galaxies is often referred to as `Fall relation' after \citet{f:83}. \citet{pfd:18} observe 92 nearby galaxies from dwarf irregulars (Im) to massive spirals (S0) excluding ellipticals and determine their Fall relation as $\log j_{\rm s} = 0.55[\log(M_{\rm s}/{\rm M_\odot})-11] + 3.34$ in the range of $7\lsim\log (M_{\rm s}/{\rm M_\odot})\lsim11.5$. \citet{mfp:19} find no evolution of this relationship to $z=1$ using the $B$- and $I$-band observations\footnote{More precisely, they use f814w and f160w bands of the Hubble Space Telescope, which roughly correspond to rest-frame $B$- and $I$-bands.}. In Fig. \ref{momenta}, the dashed line with the shaded region indicates these observations, and our results of Halo 1, 2 and 4 are consistent with the observed Fall relation. Halo 3, 5 appear to be significantly below the Fall relation. Hence, they might be considered to be elliptical galaxies in terms of the classification based on the Fall relation. Although their stellar discs are indeed relatively compact, their morphologies still appear to be discs, rather than triaxial ellipsoids\footnote{We need more detailed discussion on whether Halo 3 and 5 are elliptical or disc galaxies; it is beyond the scope of this study.}. Note that the above analysis is applied to the snapshots at $z=1$, whereas the edge-on morphologies shown in Fig. \ref{EdgeOn} are not exactly at $z=1$. We have checked and confirmed that there is no significant evolution of their morphologies to $z=1$. There could thus be tension between our simulations and the observations in the correlation between stellar masses and angular momenta, especially for low-mass galaxies. The momentum defined by Equation (\ref{AngMom}) is weighted by mass, whereas the observed Fall relation is obtained from momenta weighted by luminosities. We compute $j_{\rm s}$ weighted by luminosities in $B$- and $i$-bands by replacing $m_{{\rm s},i}$ with the luminosities of the stellar particle\footnote{$B$- and $i$-band luminosities of a stellar particle are obtained by adopting the filter functions of \citet{b:78} and \citet{s:02} to a spectral energy distribution computed with the single stellar population model. Note that we do not take into account dust attenuation.}. We find that $j_{\rm s}$ hardly depend on whether these are weighted by mass or luminosities.

\subsubsection{The Kennicutt-Schmidt law}
\begin{figure*}
  \includegraphics[bb=0 0 1659 565, width=\hsize]{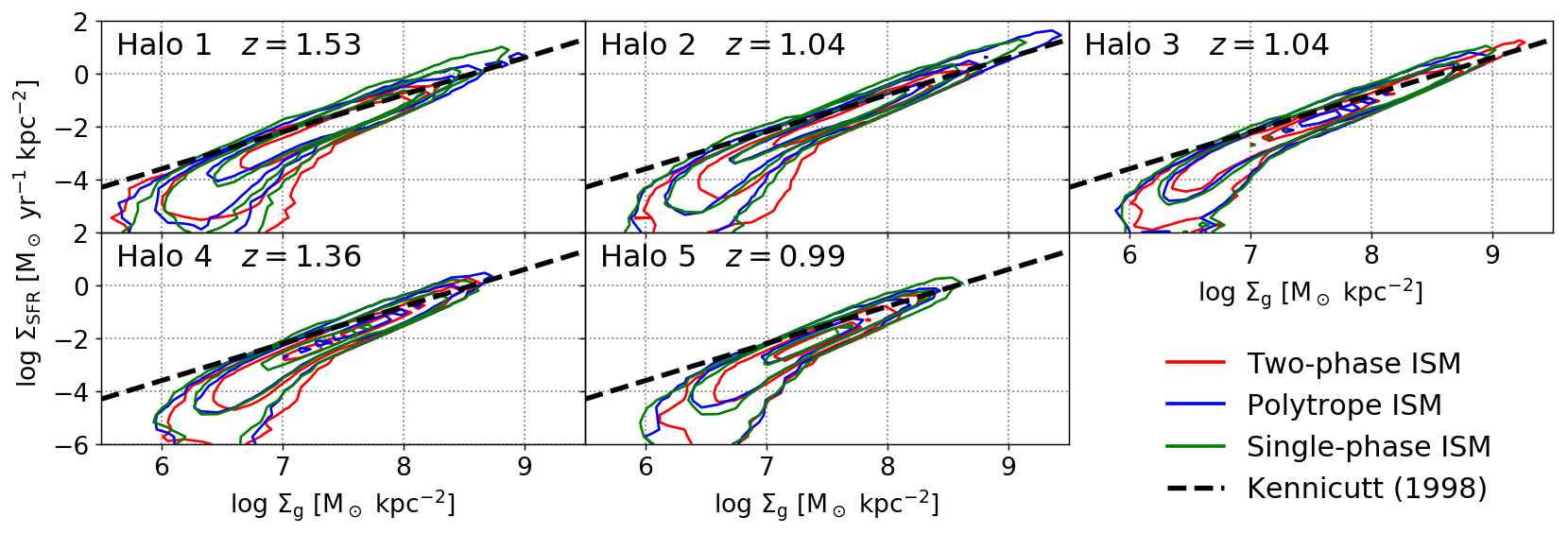}
  \caption{Scatter diagrams between $\Sigma_{\rm g}$ and $\Sigma_{\rm SFR}$ in the snapshots in Figs. \ref{snapshots1} and \ref{snapshots2}. The face-on distributions of masses and SFRs of gas cells are pixelised into areas of $0.5\times0.5~{\rm kpc^2}$, and then data points of the pixels are binned into grids whose size is $(\Delta\log\Sigma_{\rm g},\Delta\log\Sigma_{\rm SFR})=(0.13,0.27)$ in each panel of this figure. The contours delineate levels of 1, 10 and 100 data points per grid, computed by applying Gaussian kernel density estimation using Scott's rule \citep{scott:15}. The black dashed lines indicate the K-S law obtained by \citet{k:98}: $\log\Sigma_{\rm SFR}=1.4\log\Sigma_{\rm g}-12$.}
  \label{KS}
\end{figure*}
It is known that, in the Kennicutt-Schmidt (K-S) law, i.e. $\Sigma_{\rm SFR}\propto\Sigma_{\rm g}^\alpha$, the exponent $\alpha$ depends on EOS of star-forming gas even if the Schmidt law, i.e. $\rho_{\rm SFR}\propto\rho_{\rm g}^\beta$, is the same \citep{sd:08}. Accordingly, the relationship between $\Sigma_{\rm SFR}$ and $\Sigma_{\rm g}$ can depend on the ISM models assumed in our simulations. Using the snapshots in Figs. \ref{snapshots1} and \ref{snapshots2}, we pixelise the face-on distributions of gas masses and SFRs by binning them into areas of $0.5\times0.5~{\rm kpc^2}$ with the Cloud-in-Cell algorithm. Fig. \ref{KS} shows scatter diagrams between $\Sigma_{\rm SFR}$ and $\Sigma_{\rm g}$ of the pixels. In our results, no difference in the $\Sigma_{\rm g}$-$\Sigma_{\rm SFR}$ relation is seen between the runs with the three ISM models, and the relationship is consistent between all galaxies. 

Star formation in disc regions is expected to take place in two-dimensional regime, whereas giant clumps may form stars by three-dimensional collapse of gas if the clump is embedded within a gas disc. If this is the case, regions inside clumps can deviate from the K-S law of disc regions \citep[e.g.][]{eh:15}. In Fig. \ref{KS}, however, the regions inside the giant clumps in Halo 1 and 2 with the polytrope and the single-phase models --- where $\Sigma_{\rm g}$ and $\Sigma_{\rm SFR}$ are the highest --- indicate no significant deviation from the K-S law in the other regions. The dashed line indicates the observed relationship of \citet{k:98}, and all of our simulations are in agreement with the observations in the range of $\log[\Sigma_{\rm g}/({\rm M_\odot~kpc^{-2}})]\gsim7$. Thus, the relationship between $\Sigma_{\rm SFR}$ and $\Sigma_{\rm g}$ are robust with respect to the ISM models in our simulations. 

In all runs, the $\Sigma_{\rm g}$-$\Sigma_{\rm SFR}$ relations show significant offsets from the K-S law in the low-density regions with $\log[\Sigma_{\rm g}/({\rm M_\odot~kpc^{-2}})]\lsim7$. Most of the regions with such low $\Sigma_{\rm g}$ correspond to outside the gas discs in the simulations (see the top panels of Figs. \ref{snapshots1} and \ref{snapshots2}), where local gas densities are typically below the star-formation threshold $\rho_{\rm th}$. Therefore, spatial distribution of SFRs becomes discontinuous, and the $\Sigma_{\rm g}$-$\Sigma_{\rm SFR}$ relations fall off from the K-S law. \citet{sd:08} present detailed discussion about the fall-off and its dependence on $\rho_{\rm th}$. The fall-off from the K-S law at $\log[\Sigma_{\rm g}/({\rm M_\odot~kpc^{-2}})]\sim7$ is also observed in nearby galaxies \citep[][and references therein]{ke:12}. This implies that the overall effect caused by the star-formation threshold may be actually physical.

\section{Discussion}
\label{discussion}
\subsection{Influence of the ISM models and a potential problem of current cosmological simulations}
\label{potentialproblem}
We have shown that the EOS of dense gas can significantly influence clumpiness of disc galaxies in cosmological simulations. The polytrope and the single-phase ISM models can make relatively massive galaxies highly clumpy. We find, on the other hand, that no clumps with $M_{\rm clump}>10^{5.5}~{\rm M_\odot}$ form if we use the two-phase model. \cite{bdm:18} demonstrate that, in their NIHAO simulations, clustering signal of newly formed stars depends on a density threshold of star formation, and a higher density threshold leads to strong clustering. They argue that the criterion $n_{\rm H,th}\sim10~{\rm cm^3}$ is required to match observations of a local spiral galaxy. We rerun our simulations using the two-phase model with $n_{\rm H,th}=1.3$ and $13~{\rm cm^3}$ while $\rho_{\rm EOS}$ is fixed. However, we find no clump formation even in these runs. Thus, clump formation within disc regions seldom occurs at redshifts $z\gsim1$ in our cosmological simulations using the two-phase ISM model.

We find that the global properties such as the total stellar and gas masses, SFRs and metallicities of the simulated galaxies are not significantly affected by the ISM models, in spite of the significant difference of clumpiness. In addition, specific angular momenta of stellar discs are almost unchanged by switching the ISM models. Using OWLS simulations with different indices of $n=1, 4/3$ and $5/3$ for the polytrope ISM model, \citet{hsb:13} argue that these global properties as functions of stellar mass do not significantly depend on $n$. Although giant clumps are not formed in their simulation even in the case of $n=1$, they demonstrate that density contrast between spiral arms and inter-arm regions becomes sharp as $n$ decreases.

These results mentioned above can be indicative of a potential problem of the current cosmological simulations. Free parameters such as strength of stellar feedback are usually calibrated such that the global properties of simulated galaxies match statistically with observations. However, a few prominent properties such as clumpiness are not considered in the calibration. This renders the adopted models to be inaccurate or at least uncertain. Our study here shows that changing ISM models in simulations can dramatically change galactic clumpiness without significantly affecting the global properties. This implies that, even if an inaccurate ISM model is used, a cosmological simulation calibrated using the global properties is expected to hold the matching with the observations. However, such a simulation with the inaccurate ISM model cannot reproduce clumpiness correctly. Because changing ISM models hardly influences the agreement with observed global properties, the overall agreement in the global properties cannot determine which ISM model should be used. However, clumpiness of simulated galaxies can largely depend on the ISM models. ISM models used in simulations are different between studies (see Sections from \ref{twophasemodel} to \ref{singlephasemodel}). In some of large simulation projects, their datasets are publicly available and used in a number of studies. It should be kept in mind that morphologies of simulated galaxies, such as clumpiness, can significantly depend on ISM models assumed in simulations. 

From the discussion above, we suggest that ISM models, i.e. EOS of dense gas, should also be calibrated with observations of galactic morphologies such as number fractions of clumpy galaxies and mass distributions of clumps in galaxies. This kind of observations have already been available in previous studies \citep{tkt:13II,mkt:14,gfb:14,sok:16,r:17}. \citet{bmo:17} show that evolution of clumpy fractions in their large sample of zoom-in simulations using the single-phase ISM model are consistent with observations although their clumps are less bound and short-lived structures.

\subsection{Impact of clump formation on galactic properties}
\label{impact}
By switching the ISM models, we can compare clumpy and non-clumpy galaxies without changing feedback or initial conditions (i.e. mass-accretion history). From this comparison, we investigate the impact by giant clump formation on properties of the galaxies.

From our results, we find that the clump formation seen in the runs of Halo 1 and 2 does not appear to affect the galactic properties. Although giant clumps generally have intense star formation, we find that the clump formation does not largely increase the total SFRs within the galaxies; the enhancement is no more than by a factor of $\sim2$--$3$ in our simulations. This may be consistent with the observations of \citet{ggf:11}, in which they estimate the total SFR within giant clumps in a galaxy at $z\sim2$ to be nearly fifty per cent of that within the entire galaxy. Moreover, gas consumption and outflows by star formation within the clumps do not significantly decrease gas fractions or increase stellar masses of the galaxies in our simulations. \citet{sok:16} observe the redshift-evolution of number fractions of clumpy galaxies and show that it intriguingly correlates with the cosmic SFR densities; both indicate peaks at $z\simeq2$--$3$. As one of the possibilities, they argue that the increase of the cosmic SFR density can be explained as enhancement of SFRs driven by giant clump formation in disc galaxies. We infer, however, that this would not be the case since the SFRs are not significantly different between the clumpy galaxies with the single-phase model and the non-clumpy ones with the two-phase model. Moreover, clump formation does not significantly affect time-evolution of stellar masses, SFRs or gas fractions. Now that mass-accretion history of the galaxies is also the same, the clumpy and non-clumpy galaxies are expected to take similar evolution. Therefore, clump formation could not be a quenching mechanism of disc galaxies.


\section{Conclusions and summary}
\label{conclusions}
We study influence of the ISM models assumed for dense gas on galactic properties in our cosmological zoom-in simulations while keeping initial conditions and the other simulation settings unchanged. Our findings are listed below.
\begin{itemize}
  \item Clumpiness of galactic discs strongly depends on EOS of dense gas. Softer EOS can lead disc galaxies to clumpy phase at redshifts $z=1$--$2$, especially for relatively massive and extended galaxies. If harder EOS are assumed, disc galaxies do not form massive clumps even in highly gas-rich states.
  \item Global properties of the galaxies such as the total stellar masses, gas fractions, SFRs, metallicities, stellar angular momenta and $\Sigma_{\rm g}$-$\Sigma_{\rm SFR}$ relations are not significantly affected by changing the ISM models.
  \item Formation of giant clumps does not significantly alter the properties of the simulated galaxies.
\end{itemize}

A number of studies using cosmological simulations have discussed impact of stellar feedback that can reduce the number, masses and lifetimes of giant clumps. We propose, however, that EOS of dense gas is another key factor to largely affect clumpy states of galaxies in cosmological simulations. Morphologies, especially clumpiness, of simulated galaxies can strongly depend on the ISM models, and the two-phase ISM model strongly suppresses formation of giant clumps. Ones should beware of this point when studying morphologies and clumps of galaxies using simulations.

\section*{Acknowledgements}
We are grateful to the anonymous reviewer for his/her careful reading and helpful comments. We thank Volker Springel for kindly providing the simulation code {\sc Arepo} and Taira Oogi for his fascinating discussion and suggestion. This study was supported by World Premier International Research Center Initiative (WPI), MEXT, Japan and by SPPEXA through JST CREST JPMHCR1414. SI receives the funding from KAKENHI Grant-in-Aid for Young Scientists (B), No. 17K17677. The numerical computations presented in this paper were carried out on Cray XC50 at Center for Computational Astrophysics, National Astronomical Observatory of Japan.


\appendix

\section{Clump identification scheme}
\label{clumpdetection}
Our clump identification scheme is basically the same as those used in \citet{mdc:13,mdc:17} and \citet{idm:16}. First, we apply a spherical Gaussian smoothing with a narrow full width at half maximum (FWHM) of $0.33~{\rm kpc}$ to the three-dimensional gas density field, and we deposit masses in uniform grids whose size is $0.1~{\rm kpc}$. This procedure washes out noise at the resolution level of the simulations but retains the desired clumpy structures. Then, we apply a wide Gaussian smoothing with an FWHM of $2.5~{\rm kpc}$ to the original gas density field. The wide Gaussian filter smooths out large-scale structures within a disc. Then, we obtain the residual: $\delta\equiv(\rho_{\rm N}-\rho_{\rm W})/\rho_{\rm W}$, where $\rho_{\rm N}$ and $\rho_{\rm W}$ are the gas densities smoothed with the narrow and wide Gaussian filters. Next, we group neighbouring grids having $\delta>10$ into a clump, and stellar particles in a clump are assigned to the grouped grids that contain the stars. Positions of the identified clumps are defined as mass-weighted centres of the grouped gas and stars. We exclude clumps within the central region of $R<3~{\rm kpc}$ to avoid bulges. Moreover, we ignore clumps in the regions above the height of $|z|=5~{\rm kpc}$ from disc planes and outside the regions depicted in Figs \ref{snapshots1} and \ref{snapshots2} since they are not formed within the discs but accreting from outer haloes. 

Next, we compute a spherically averaged mass distribution $m_{\rm clump}$ of the identified clump, where $m_{\rm clump}$ is mass of baryon and DM enclosed within a distance from the clump centre. Similarly, we obtain the total mass distribution $M_{\rm tot}(r)$ of the whole system, where $r$ is a galactocentric distance. Assuming the identified clump to be in a circular orbit, we can estimate a tidal radius $r_{\rm t}$ of the clump as
\begin{equation}
r_{\rm t}=r_{\rm clump}\left[\frac{m_{\rm clump}(r_{\rm t})/M_{\rm tot}(r_{\rm clump})}{3-\frac{\mathrm{d}\ln M_{\rm tot}}{\mathrm{d}\ln r}|_{r_{\rm clump}}}\right]^{\frac{1}{3}},
\label{tidalradius}
\end{equation}
where $r_{\rm clump}$ is a galactocentric distance of the clump centre. We can estimate escape velocities as a function of distances from the clump centre by defining the gravitational potential of the clump to be $\Phi_{\rm clump}|_{r_{\rm t}}=0$. We remove matters outside $r_{\rm t}$ or whose velocities are higher than their escape velocities. We finally compute the total masses of baryons $M_{\rm clump}$ and DM $M_{\rm DM}$ that are expected to be bound to the clump. Accreting dwarf galaxies can be mixed in with giant clumps in disc regions although they are only a few. Because such `ex-situ' clumps generally have large amounts of bound DM \citep{mdc:13}, we exclude clumps that indicate $M_{\rm DM}/(M_{\rm clump}+M_{\rm DM})>0.1$. We define the total mass of bound baryon $M_{\rm clump}$ to be the mass of the clump.

\end{document}